\documentclass[11pt]{article}
\usepackage{amsmath,amssymb,cite,graphicx,color}
\usepackage[margin=1.2in]{geometry}

\def\be{\begin{equation}}
\def\ee{\end{equation}}
\def\bea{\begin{eqnarray}}
\def\eea{\end{eqnarray}}
\def\eq#1{(\ref{#1})}

\def\chiv{\eta}
\def\cG{{\cal G}}
\def\cH{{\cal H}}
\def\cM{{\cal M}}
\def\cO{{\cal O}}
\def\cR{{\cal R}}
\def\prd{Phys.\ Rev.\ D}
\def\plb{Phys.\ Lett.\ B}

\begin{document}

\centerline{\bf \large RG flow of Weyl--invariant dilaton gravity}
\bigskip
\centerline{R. Percacci}
\medskip
\centerline{SISSA, via Bonomea 265, 34136 Trieste, Italy}
\centerline{and INFN, Sezione di Trieste, Italy}

\begin{abstract}
Any theory can be made Weyl invariant by introducing a dilaton.
It is shown how to construct renormalization group equations
for gravity that maintain this property.
Explicit calculations are given only in the simplest approximation,
namely for the one loop beta functions of a dilaton 
conformally coupled to a dynamical metric,
but the results have wider validity.
This formalism could be used to define the meaning of a theory
with a position-dependent cutoff: it is equivalent to
a theory with a constant cutoff, but a conformally related metric.
\end{abstract}


\section{Introduction}


\subsection{Scalar-tensor theories}

This work has two main motivations.
The first is to fill a gap in the literature concerning
the renormalization group flow of scalar-tensor theories.
This problem has been studied earlier using functional renormalization group methods
\cite{griguolo,perini2,narain}, with the aim of calculating the gravitational contributions to
the scalar beta functions and investigating the existence of a fixed point (FP) for
a system with Lagrangians of the general form $(\nabla\phi)^2+F(R,\phi^2)$,
where the function $F$ is usually polynomial in both arguments.
It was found that in four dimensions the system has a ``Gaussian matter FP'',
of the form $F_*(R,0)$, {\it i.e.} at the FP the scalar is only minimally coupled to gravity
\cite{perini2,narain}.
A question that has been raised several times is the behavior of
the nonminimal coupling $\xi \phi^2 R$, which is conformal for $\xi=1/6$
and is expected to have a FP there.
This is not the case in the calculations mentioned above,
but the reason is simple: a mass-dependent cutoff was used, breaking Weyl (conformal) 
invariance even if present in the original Lagrangian.
It is therefore not surprising that in those calculations
nothing special seems to happens at $\xi=1/6$.
The first motivation of this paper is to investigate the existence of a Weyl 
covariant cutoff procedure such that $\xi=1/6$ is a FP.

Such a procedure would have further benefits.
The class of theories that we shall consider can be seen as the result of
making an arbitrary theory Weyl invariant by
introducing a dilaton field $\chi$ and replacing all dimensionful couplings $g_i$
by dimensionless couplings $\hat g_i$ multiplied by appropriate powers of the dilaton.
It is interesting to compare the renormalization of the original
dimensionful couplings $g_i$ to the renormalization of their dimensionless cousins $\hat g_i$.
We shall see that the equivalence between the original theory and its Weyl invariant 
version can be maintained along the RG flow,
and the results obtained earlier regarding the existence 
of a gravitational FP will be confirmed, but with some interesting quirks.
In order not to burden the discussion with unnecessary complications
we will restrict ourselves to the simplest approximation
to the simplest truncation of the action,
namely the one loop beta functions of a dilaton conformally coupled to the metric.
This corresponds to the ``Einstein-Hilbert truncation'' 
of pure gravity \cite{souma,lauscher,litim1}.
It will be clear that the same techniques can be applied also
to more complicated truncations discussed in
\cite{lauscher2,codello1,cpr1,cpr2,bms,niedermaier,ghost,immirzi}
and to theories with matter.

\subsection{Non-constant cutoffs}

The second motivation is the desire to define a theory also in the presence 
of a cutoff that depends on position.
In particle physics experiments one usually considers a small region of
spacetime and sets up the apparatus in such a way as to have as much control
as possible over everything that goes in and out of it.
The process is characterized by a momentum scale that is a function of the momenta 
of the in- and outgoing particles, and the description of the process
is optimized (in the sense that radiative corrections are minimized)
by choosing a renormalization point that coincides with this scale.
In the Wilsonian description of the theory that we will use here,
the choice of renormalization point corresponds to a choice of an infrared cutoff.
The chosen scale is a global property of the experiment and there is no reason
to think that it could change with position.
When we think of applying quantum field theory to astrophysical systems 
or to cosmology, we encounter very different situations.
The typical energy scales involved in physical processes
may change very strongly with time or with position,
and a position-dependent cutoff may be desirable.

The way in which this has been mostly handled so far is by identifying the cutoff
with a specific function of position (usually via some prescribed function of the fields) 
and then to make the couplings depend on position via this cutoff function. 
In cosmology, different choices of cutoff have been considered 
in the past, for example $1/a(t)$ \cite{florean,bauer}, 
$1/t$ \cite{cosmo1,rw1,rw2} or the Hubble parameter $H=\dot a/a$
\cite{entropy,bcp}. See also \cite{bab,sha,weinstein}.
Replacing the classical, constant couplings by their running counterparts 
at the level of the equations of motion, leads to apparent violation
of the conservation of the energy momentum tensor, 
which may be interpreted as a kind of diffeomorphism anomaly.
The simplest example of this phenomenon occurs when one turns Newton's
constant and the cosmological constant in Einstein's equations 
$G^{\mu\nu}+\Lambda g^{\mu\nu}=8\pi GT^{\mu\nu}$
into functions of position. Due to the contracted Bianchi identity,
taking the divergence of both sides gives 
\be
\label{anom}
\nabla_\mu T^{\mu\nu}=-\frac{\nabla_\mu G}{G}\, T^{\mu\nu}+\frac{1}{8\pi G}\nabla^\nu\Lambda .
\ee
If the energy-momentum tensor derives from the action of a massless scalar field,
it is going to be conserved when one uses the scalar equation of motion.
Then, the r.h.s. of equation \eq{anom} will vanish {\it on shell}.
If the matter field was massive or had other interactions,
equation \eq{anom} will contain also other terms involving the
derivatives of these additional Lagrangian parameters.
In any case, equation \eq{anom} in itself does not signal an inconsistency
of the theory, but rather gives conditions that will be obeyed on shell
by the energy-momentum tensor.
\footnote{This parallels the following well-known fact:
in an anomalous gauge theory, the anomaly vanishes
on shell and therefore is not inconsistent with the equations of motion 
of the theory, see e.g. \cite{rajaraman}.}

On the other hand, if the energy-momentum tensor describes a phenomenological
fluid, there is no a priori reason for it to be conserved.
One could then argue that energy-momentum is conserved at the microscopic level
and that \eq{anom} is a measure of the energy and momentum of the field modes
that are removed from the description of the system by coarse graining.
Again, it should not be seen as a pathology.
In fact, it can be exploited to account for all the entropy 
that is observed in the universe \cite{entropy}.

Still, in some applications it would be desirable to have a formalism
that is capable of dealing with non-constant cutoffs,
while completely avoiding the preceding issues.
As we noted above, in theories with a dilaton, dimensionful couplings
are effectively replaced by fields, and therefore can become position-dependent.
This suggests that making the theory Weyl-invariant by introducing a dilaton
may be a way of achieving this goal.

\subsection{The dilaton and Weyl invariance}

We will assume that special relativity and quantum mechanics hold true,
so without any further thought we can choose units where $c=1$, $\hbar=1$.
Then, everything has dimension of a power of length.
The length dimension of a physical quantity $q$ is the number $d$
which enters in the transformation rule $q\mapsto \Omega^d q$,
under the action of the multiplicative group of the positive real numbers.
Such transformations will be called {\it global rescalings}.
In quantum field theory in Minkowski space, global rescalings are usually
viewed as spacetime transformations, acting as $x^\mu\to\Omega x^\mu$;
this is consistent with the use of coordinates having dimension of length.
In a gravitational context
it seems more convenient to view all coordinates are mere numerical labels 
for the points of spacetime, carrying no dimension.
Then, the transformation of the fields do not involve any change in their argument.
We will stick to this convention throughout this paper. 
There follows from the definition of the line element, 
that the metric tensor has dimension of area. This means that it transforms as
\be
\label{conftransf}
g_{\mu\nu}\mapsto \Omega^2 g_{\mu\nu}\ .
\ee
The dimensions of the fields $\psi_a$ and couplings $g_i$,
called $d_a$ and $d_i$ respectively,
are determined by requiring that the action $S(g_{\mu\nu},\psi_a,g_i)$
is dimensionless, {\it i.e.} that
\be
\label{weylinv}
S(g_{\mu\nu},\psi_a,g_i)=S(\Omega^2 g_{\mu\nu},\Omega^{d_a}\psi_a,\Omega^{d_i}g_i)\ .
\ee
It does not matter here whether the metric is fixed or dynamical.
One finds that the dimensions of scalar, spinor and vector fields
are $1$, $3/2$ and $0$, respectively. One can easily convince oneself that
the dimensions of all parameters in the Lagrangian, such as masses and couplings, 
are the same as in the more familiar case when coordinates have 
dimension of length.

Every action is invariant under these transformations, by construction.
This is just the statement of dimensional analysis.
Sometimes the action is also invariant under the infinite dimensional
abelian group of maps from spacetime into the positive reals.
These transformations look the same as above, but with $\Omega$ 
now regarded as a function of $x$.
Such transformations are called {\it  Weyl transformations}.
Since Lagrangian parameters must be independent of position,
a necessary condition for Weyl invariance is that all parameters
in the Lagrangian be dimensionless.

So far we have discussed rescalings of {\it all} dimensionful 
quantities appearing in a theory. These transformations
should not be confused with another class of transformations
called (global or local) {\it scale transformations},
which form the same abstract group but act only on
the fields, not on the parameters of the theory
(for discussions, see for example \cite{bransdicke,bekenstein,coleman}).
The reason for considering these transformations is that theories 
with different masses or couplings are viewed as different theories.
Then, one says that a theory is (globally or locally) scale invariant if 
\be
\label{scaleinv}
S(g_{\mu\nu},\psi_a,g_i)=S(\Omega^2 g_{\mu\nu},\Omega^{d_a}\psi_a,g_i)\ .
\ee
Comparing \eq{weylinv} and \eq{scaleinv} we see that a theory is globally
scale invariant if and only if $d_i=0$ for all $i$.
Since in a Weyl invariant theory all parameters $g_i$ are dimensionless,
Weyl invariance is equivalent to local scale invariance.

When one uses dimensionful coordinates, a distinction has to be made 
between the canonical dimension and the Weyl or scaling dimension of a field.
For example, when coordinates have dimension of length, the canonical dimension
of a vector field is mass but its Weyl dimension is zero.
One advantage of using dimensionless coordinates is that
the two notions are the same.
In quantum field theory, canonically normalized fields acquire noninteger
(anomalous) scaling dimensions \cite{coleman}.
Equivalently, one can work with fields whose scaling dimension is not
affected by quantum corrections but then the kinetic terms contain
dimensionless wave function renormalization constants $Z_a$
which depend on the renormalization scale.
We will work with these non-canonically normalized fields.

Normally one works with units of length and mass that are implicitly
assumed to be constant over space and time,
but in Weyl's theory such a notion is declared physically meaningless.
The unit must be allowed to vary arbitrarily from point to point.
It can thus be viewed as a scalar field $\chi$ of dimension mass,
which can be written as $\chi(x)=e^{\sigma(x)} \chi_0$.
It is important that $\chi$ is not allowed to be zero anywhere,
otherwise it would not be a unit anymore.
The field $\chi$ will be called the dilaton.

We can start from any action and rewrite it in a way that is manifestly
invariant under Weyl transformations, by making the presence 
of the unit/dilaton explicit.
We begin by observing that the ordinary covariant derivative of a field is not a Weyl-covariant notion:
under Weyl transformations, terms involving derivatives of the
parameter $\Omega$ appear. Such terms can be removed by defining
a Weyl covariant derivative as follows.
With the dilaton we construct a pure-gauge abelian gauge field
$\kappa_\mu=-\chi^{-1}\partial_\mu\chi$, transforming under \eq{conftransf} as
$\kappa_\mu\mapsto \kappa_\mu+\Omega^{-1}\partial_\mu\Omega$.
Let $\nabla_\mu$ be the covariant derivative with respect to the
Levi-Civita connection of the metric $g$.
Define a new (non-metric) connection
\be
\hat\Gamma_\mu{}^\lambda{}_\nu=
\Gamma_\mu{}^\lambda{}_\nu-
\delta^\lambda_\mu\kappa_\nu-\delta^\lambda_\nu\kappa_\mu
+g_{\mu\nu}\kappa^\lambda\ ,
\ee
where $\Gamma_\mu{}^\lambda{}_\nu$ are the Christoffel symbols of $g$.
The corresponding covariant derivative is denoted $\hat\nabla$.
The connection coefficients $\hat\Gamma$ are invariant under \eq{conftransf}.
For any tensor $t$ of length dimension $w$ 
define the covariant derivative $Dt$ to be
\be
\label{wcovder}
D_\mu t=\hat\nabla_\mu t-w\kappa_\mu t\ ,
\ee
where all indices have been suppressed.
The curvature of $D$ is defined by
\be
[D_\mu,D_\nu]v^\rho=\cR_{\mu\nu}{}^\rho{}_\sigma v^\sigma\ .
\ee
The tensor $\cR_{\mu\nu}{}^\rho{}_\sigma$ is Weyl invariant,
and raising and lowering indices one obtains Weyl covariant expressions
of different dimensions.
A direct calculation gives the explicit expression
\bea
\label{wcovcurv}
\cR_{\mu\nu\rho\sigma}&=& R_{\mu\nu\rho\sigma}
+g_{\mu\rho}\left(\nabla_\nu\kappa_\sigma+\kappa_\nu\kappa_\sigma\right)
-g_{\mu\sigma}\left(\nabla_\nu\kappa_\rho+\kappa_\nu\kappa_\rho\right)
\nonumber
\\
&&
\!\!\!\!\!\!\!\!\!\!\!\!\!\!\!\!\!\!
-g_{\nu\rho}\left(\nabla_\mu\kappa_\sigma+\kappa_\mu\kappa_\sigma\right)
+g_{\nu\sigma}\left(\nabla_\mu\kappa_\rho+\kappa_\mu\kappa_\rho\right)
-\left(g_{\mu\rho}g_{\nu\sigma}-g_{\mu\sigma}g_{\nu\rho}\right)\kappa^2\ .
\eea

Now start from a generic action for matter and gravity of the form
$S(g_{\mu\nu},\psi_a,g_i)$.
Express every parameter $g_i$ as $g_i=\chi^{d_i} \hat g_i$,
where $\hat g_i$ is dimensionless.
Replace all covariant derivatives $\nabla$ by Weyl covariant derivatives $D$ and all curvatures $R$ by the Weyl covariant curvatures $\cR$.
Now all the terms appearing in the action are products of Weyl covariant
objects, and local Weyl invariance just follows from the fact that the action
is dimensionless.
In this way we have defined an action $\hat S(g_{\mu\nu},\chi,\psi_a,\hat g_i)$.
It contains only dimensionless couplings $\hat g_i$, 
and is Weyl invariant by construction.

Because the dilaton transforms as $\chi\to\Omega^{-1}\chi$
with $\Omega$ a nowhere zero function, we can transforms any
dilaton configuration into any other.
In particular we can set $\sigma=0$ or $\chi=\chi_0$.
This brings us back to the situation in which the unit can be seen
as constant, but now it is a gauge choice.
If we make this choice, all derivatives of $\chi$ in the action vanish,
and the factors of $\chi_0$ combine with the $\hat g_i$
to reconstruct the dimensionful $g_i$, giving back the original action $S$:
\be
\hat S(g_{\mu\nu},\chi_0,\psi_a,\hat g_i)=S(g_{\mu\nu},\psi_a,g_i)\ .
\ee

In conclusion, there is 
a one-to-one correspondence between the class of all
theories depending on a metric and fields $\psi_a$,
and Weyl invariant theories depending on the same fields plus a dilaton.
Every theory is invariant under global rescalings because
dimensional analysis always holds true, and it can be made manifestly
Weyl invariant by the above covariantization procedure.
We shall be interested in theories that are Weyl invariant in this sense.


\section{The action and equations of motion}


We will consider Euclidean action functionals of the general form
\be
S =  \int d^4 x \sqrt{g}\left[\lambda Z^2\chi^4
-\frac{1}{2}Z
\left(\xi\chi^2 R+g^{\mu\nu}\partial_\mu\chi\partial_\nu\chi\right)
\right]
\ , 
\label{action}
\ee
containing three dimensionless couplings $Z$, $\xi$ and $\lambda$.
It is clear that $Z$ is redundant: it could be set to one by a simple rescaling
of $\chi$. Still, it plays a role in the renormalization of the theory
so it is useful to keep it for future reference. We will return to this point
in section 6.1. 

The action \eq{action} is invariant under global rescalings \eq{conftransf}
together with $\chi\mapsto \Omega^{-1}\chi$,
with $\Omega$ a constant and the couplings held fixed. 
Under \eq{conftransf} the curvature scalar transforms as
$R\to \Omega^{-2}\left(R-6\Omega^{-1}\nabla^2\Omega\right)$,
so one sees that if $\xi=1/6$, \eq{action} is also invariant under local rescalings, i.e. transformations \eq{conftransf} with $\Omega$
a positive function on spacetime.
In this case the action can also be written in the form
\be
\label{confaction}
S =  \int d^4 x \sqrt{g}\left[\lambda Z^2\chi^4
-\frac{1}{12}Z\chi^2\cR
\right]
\ , 
\ee
where the Weyl-covariant curvature $\cR$ is defined as in \eq{wcovcurv}.
When $\xi=1/6$ the field $\chi$ is unphysical and if we set
\be
\label{conversion}
Z\chi^2=\frac{12}{16\pi G}\ ;
\qquad
\lambda=\frac{2\pi}{9}G\Lambda
\ee
the action \eq{confaction} is equivalent to the Hilbert action
\be
\label{hilbert}
S =  \int d^4 x \sqrt{g} \frac{1}{16\pi G}(2\Lambda-R) 
\ ,
\ee

This equivalence can be seen also at the level of the equations of motion (EOMs).
The first variation of \eq{action} yields
\be
\delta S=Z\int d^4 x\sqrt{g}
\left[\delta g_{\mu\nu}E^{\mu\nu}+\delta\chi E\right]\ ,
\ee
where
\begin{align}
E_{\mu\nu}&=\frac{1}{2}\xi\chi^2\left(R_{\mu\nu}-\frac{1}{2}g_{\mu\nu}R\right)+
\left(\frac{1}{2}-\xi\right)\nabla_\mu\chi\nabla_\nu\chi
+\left(\xi-\frac{1}{4}\right)g_{\mu\nu}(\nabla\chi)^2
\nonumber
\\
&
-\xi \chi\nabla_\mu\nabla_\nu\chi
+\xi g_{\mu\nu}\chi\nabla^2\chi
+\frac{1}{2}\lambda Z\chi^4 g_{\mu\nu}
\end{align}
and
\be
E=\nabla^2\chi-\xi R\chi+4\lambda Z\chi^3\ .
\ee
Thus the EOM of the theory are
\be
E_{\mu\nu}=0\ ;\qquad E=0\ .
\ee
Taking the trace of the tensor EOM we get
\be
0=-\frac{1}{2}\xi\chi^2 R+\left(3\xi-\frac{1}{2}\right)(\nabla\chi)^2
+3\xi\chi\nabla^2\chi+2\lambda Z\chi^4\ .
\ee
If $\xi=1/6$ this is equivalent to the scalar EOM.


\section{Expansion of the action}


For the quantum theory we need the second variation of the action.
From here on we will use the symbols $g_{\mu\nu}$ and $\chi$ to denote
background fields and $h_{\mu\nu}=\delta g_{\mu\nu}$ and $\chiv=\delta\chi$
for the variations.
Expanding to second order in powers of $h_{\mu\nu}$ and $\chiv$, 
and discarding total derivative terms, the quadratic part of the action is given by
\begin{align}
S^{(2)} =&  \frac{1}{2}Z \int d^4 x \sqrt {g}
\Biggl\{\frac{1}{2}\xi\chi^2\Biggl[
-\frac{1}{2} h_{\mu\nu}\nabla^2 h^{\mu\nu}
+h_{\mu\nu} \nabla^\mu\nabla_\rho h^{\rho\nu}
- h \nabla^\mu\nabla^\nu h_{\mu\nu}
+ \frac{1}{2}h\nabla^2 h
\nonumber
\\
&
-h_{\mu\nu}R^{\mu\rho\nu\sigma}h_{\rho\sigma}
-h_{\mu\nu}R^{\nu\sigma}h^\mu{}_\sigma
+h R^{\rho\sigma}h_{\rho\sigma}
+\frac{1}{2}\left(R-\frac{2}{\xi}\lambda Z\chi^2\right)
\left(h_{\mu\nu}h^{\mu\nu}-\frac{1}{2}h^2\right)\Biggr]
\nonumber
\\
&
+\left(\frac{1}{2}-\frac{3}{2}\xi\right)\nabla^\rho\chi\nabla^\sigma\chi h h_{\rho\sigma}
+(2\xi-1)h_{\mu\nu}\nabla^\nu\chi\nabla^\sigma\chi h^\mu{}_\sigma
+2\xi \chi\nabla^\nu\nabla^\sigma\chi h_{\mu\nu}h^\mu{}_\sigma
\nonumber
\\
&
-\frac{3}{2}\xi\chi \nabla^\mu\nabla^\nu\chi h h_{\mu\nu}
+\left(\frac{1}{4}-\frac{3}{4}\xi\right)(\nabla\chi)^2 h_{\mu\nu}h^{\mu\nu}
+\left(\frac{1}{4}\xi-\frac{1}{8}\right)(\nabla\chi)^2 h^2
\nonumber
\\
&
-\frac{3}{4}\xi \chi\nabla^2\chi h_{\mu\nu}h^{\mu\nu}
+\frac{1}{4}\xi\chi\nabla^2\chi h^2
+\xi \chi\nabla_\lambda\chi \left(h_{\mu\nu}\nabla^\mu h^{\lambda\nu}
- h\nabla_\rho h^{\rho\lambda}\right)
\nonumber
\\
&
+\chiv\Bigl[-2\xi\chi\nabla^\mu\nabla^\nu h_{\mu\nu}
+2\xi\chi\nabla^2 h
-2\nabla^\rho\chi\nabla^\sigma h_{\rho\sigma}
+\nabla^\rho\chi\nabla_\rho h
\nonumber
\\
&
-2 \nabla^\rho\nabla^\sigma\chi h_{\rho\sigma}
+\nabla^2\chi h
+2\xi\chi R^{\rho\sigma}h_{\rho\sigma}
-\xi\chi R h
+4\lambda Z\chi^3 h
\Bigr]
\nonumber
\\
&
+\chiv\left(\nabla^2-\xi R+12\lambda Z\chi^2\right)\chiv
\Biggr\}\ .
\label{a2}
\end{align}
We define the Hessian $\cH$ by:
\be
S^{(2)}=
\frac{1}{2}Z\cH((h,\chiv),(h,\chiv))=
\frac{1}{2}Z \int d^4 x \sqrt {g}
\left(\begin{array}{cc}\!\!h_{\mu\nu}&\chiv\!\!\end{array}\right)
\left(\begin{array}{cc}\cH_{hh}^{\mu\nu\rho\sigma}&\cH_{h\chiv}^{\mu\nu}\\
\cH_{\chiv h}^{\rho\sigma}&\cH_{\chiv\chiv}\end{array}\right)
\left(\begin{array}{c}h_{\rho\sigma}\\ \chiv\end{array}\right)\ .
\ee
Note that an irrelevant overall factor of $Z$ has been extracted.
From \eq{a2} we get after symmetrizing in the arguments:
\begin{align}
\cH_{hh}^{\mu\nu\rho\sigma}=&
\frac{1}{2}\xi\chi^2\Biggl[
-\frac{1}{2}\mathbf{1}^{\mu\nu\rho\sigma} \nabla^2 
+g^{(\nu|\sigma}\nabla^{|\mu)}\nabla^\rho
- \frac{1}{2}g^{\mu\nu}\nabla^\rho\nabla^\sigma 
-\frac{1}{2} g^{\rho\sigma}\nabla^{(\mu}\nabla^{\nu)}
+ \frac{1}{2}g^{\mu\nu}g^{\rho\sigma}\nabla^2
\nonumber
\\
&
\ \ \ \ \ \ \ \ -R^{\mu\rho\nu\sigma}
-g^{(\mu|\rho}R^{|\nu)\sigma}
+\frac{1}{2} g^{\mu\nu}R^{\rho\sigma}
+\frac{1}{2} R^{\mu\nu}g^{\rho\sigma}
+\left(R-\frac{2}{\xi}\lambda Z\chi^2\right)
K^{\mu\nu\rho\sigma}\Biggr]
\nonumber
\\
&
+\frac{1}{2}\xi \chi
\left(2g^{(\mu|\rho}\nabla^\sigma\chi \nabla^{|\nu)} 
-\mathbf{1}^{\mu\nu\rho\sigma}\nabla^\lambda\chi\nabla_\lambda
-2g^{\mu\nu}\nabla^\rho\chi\nabla^\sigma
+g^{\mu\nu}g^{\rho\sigma}\nabla^\lambda\chi\nabla_\lambda \right)
\nonumber
\\
&
+\left(\frac{1}{4}-\frac{1}{2}\xi\right)g^{\rho\sigma}\nabla^{(\mu|}\chi\nabla^{|\nu)}\chi
+\left(\frac{1}{4}-\xi\right)g^{\mu\nu}\nabla^\rho\chi\nabla^\sigma\chi
+(2\xi-1)g^{(\mu|\rho}\nabla^{|\nu)}\chi\nabla^\sigma\chi 
\nonumber
\\
&
+2\xi\chi g^{(\mu|\rho}\nabla^{|\nu)}\nabla^\sigma\chi 
-\frac{1}{2}\xi\chi g^{\rho\sigma}\nabla^\mu\nabla^\nu\chi 
-\xi\chi g^{\mu\nu}\nabla^\rho\nabla^\sigma\chi
+\left(\frac{1}{4}-\xi\right)(\nabla\chi)^2 \mathbf{1}^{\mu\nu\rho\sigma}
\nonumber
\\
&
+\left(\frac{1}{2}\xi-\frac{1}{8}\right)(\nabla\chi)^2 g^{\mu\nu}g^{\rho\sigma}
-\xi \chi\nabla^2\chi \mathbf{1}^{\mu\nu\rho\sigma}
+\frac{1}{2}\xi\chi\nabla^2\chi g^{\mu\nu}g^{\rho\sigma}
\label{exp1}
\end{align}
\begin{align}
\cH_{h\chiv}^{\mu\nu}&=
-\xi\chi\nabla^\mu\nabla^\nu 
+\xi\chi g^{\mu\nu}\nabla^2 
+\xi\chi R^{\mu\nu}
-\frac{1}{2}\xi\chi R g^{\mu\nu}
+2\lambda Z\chi^3 g^{\mu\nu}
\nonumber
\\
&
+(1-2\xi)\nabla^{(\mu}\chi\nabla^{\nu)} 
+\left(2\xi-\frac{1}{2}\right)g^{\mu\nu}\nabla^\lambda\chi\nabla_\lambda 
-\xi\nabla^\mu\nabla^\nu\chi 
+\xi\nabla^2\chi g^{\mu\nu}
\label{exp2}
\\
\cH_{\chiv h}^{\rho\sigma}&=
-\xi\chi\nabla^{\rho}\nabla^{\sigma} 
+\xi\chi g^{\rho\sigma}\nabla^2 
+\xi\chi R^{\rho\sigma}
-\frac{1}{2}\xi\chi R g^{\rho\sigma}
+2\lambda Z\chi^3 g^{\rho\sigma}
\nonumber
\\
&
-\nabla^{\rho}\chi\nabla^{\sigma} 
+\frac{1}{2}g^{\rho\sigma}\nabla^\lambda\chi\nabla_\lambda 
- \nabla^\rho\nabla^\sigma\chi 
+\frac{1}{2}\nabla^2\chi g^{\rho\sigma}
\label{exp3}
\\
\cH_{\chiv\chiv}&=
\nabla^2-\xi R
+12\lambda Z\chi^2
\label{exp4}
\end{align}
where
\be
K^{\mu\nu\rho\sigma}=\frac{1}{2}
\left(\mathbf{1}^{\mu\nu\rho\sigma}-\frac{1}{2}g^{\mu\nu}g^{\rho\sigma}\right)\ .
\ee

Symmetrization under the interchange $\rho \leftrightarrow \sigma$ 
is not indicated explicitly but has to be performed where needed. 
Of course this symmetrization is automatic when the
operators act on a symmetric tensor $h_{\rho\sigma}$.
Symmetrization under the interchange $\mu \leftrightarrow \nu$ 
is indicated explicitly.
Symmetry under the interchange of the arguments 
$\cH(\psi,\theta)=\cH(\theta,\psi)$
is not obvious and requires integrations by parts.

A check on this operator comes from the fact that
\be
\label{gaugediffeo}
\left(\begin{array}{cc}\cH_{hh}^{\mu\nu\rho\sigma}&\cH_{h\chiv}^{\mu\nu}\\
\cH_{\chiv h}^{\rho\sigma}&\cH_{\chiv\chiv}\end{array}\right)
\left(\begin{array}{c}\nabla_\rho\epsilon_\sigma+\nabla_\sigma\epsilon_\rho\\ \epsilon^\lambda\nabla_\lambda\chi\end{array}\right)=0\ ,
\ee
which follows from the diffeomorphism invariance of the action.
For $\xi=1/6$ there is the further identity
\be
\label{gaugeweyl}
\left(\begin{array}{cc}\cH_{hh}^{\mu\nu\rho\sigma}&\cH_{h\chiv}^{\mu\nu}\\
\cH_{\chiv h}^{\rho\sigma}&\cH_{\chiv\chiv}\end{array}\right)
\left(\begin{array}{c}2\omega g_{\rho\sigma}\\ -\omega\chi\end{array}\right)=0
\ee
which follows from Weyl invariance of the action.
Both identities require using the EOMs.

If $\chi$ is constant, only the first two lines
of $\cH_{hh}^{\mu\nu\rho\sigma}$ survive.
If we replace $Z\xi\chi^2/2$ by $1/(16\pi G)$, they are just the
Hessian of the Hilbert action.
Now we observe that in the special case $\xi=1/6$ all the terms involving
derivatives of $\chi$ conspire to turn ordinary covariant derivatives $\nabla$
into Weyl covariant derivatives $D$ 
\footnote{It is useful to note that the covariant derivatives 
of the backgrounds vanish: $D\chi=0$, $Dg_{\mu\nu}=0$.}, 
and curvatures $R$ into Weyl curvatures $\cR$:
\begin{align}
\cH_{hh}^{\mu\nu\rho\sigma}=&
\frac{1}{12}\chi^2\Biggl[
-\frac{1}{2}\mathbf{1}^{\mu\nu\rho\sigma} D^2 
+g^{(\nu|\sigma}D^{|\mu)}D^\rho
- \frac{1}{2}g^{\mu\nu}D^\rho D^\sigma 
-\frac{1}{2} g^{\rho\sigma}D^{(\mu}D^{\nu)}
+ \frac{1}{2}g^{\mu\nu}g^{\rho\sigma}D^2
\nonumber
\\
&
-\cR^{\mu\rho\nu\sigma}\!
-g^{(\mu|\rho}\cR^{|\nu)\sigma}
+\frac{1}{2} (g^{\mu\nu}\cR^{\rho\sigma}\!
+\cR^{\mu\nu}g^{\rho\sigma})
+\left(\cR-12\lambda Z\chi^2\right)K^{\mu\nu\rho\sigma}\Biggr]
\label{exp5}
\\
\cH_{h\chiv}^{\mu\nu}=&\cH_{\chiv h}^{\mu\nu}=
\frac{1}{6}\chi \left(g^{\mu\nu}D^2-D^\mu D^\nu 
+\cR^{\mu\nu}-\frac{1}{2} \cR g^{\mu\nu}\right)+2\lambda Z\chi^3 g^{\mu\nu}
\\
\cH_{\chiv\chiv}=&
D^2-\frac{1}{6}\cR+12\lambda Z\chi^2\ .
\label{exp8}
\end{align}
This  means that $\cH_{hh}^{\mu\nu\rho\sigma}$ is just the conformal covariantization 
of the Hessian of the Hilbert action, and the other pieces in the Hessian are also manifestly
conformally covariant, in the sense that
\begin{align}
\cH^{\mu\nu\rho\sigma}_{hh(\Omega^2 g_{\mu\nu},\Omega^{-1}\chi)}(\Omega^2 h_{\rho\sigma})
=&\ \Omega^{-6} \cH^{\mu\nu\rho\sigma}_{hh(g_{\mu\nu},\chi)}h_{\rho\sigma}\ ,
\\
\cH^{\mu\nu}_{h\chiv(\Omega^2 g_{\mu\nu},\Omega^{-1}\chi)}(\Omega^{-1}\chiv)
=&\ \Omega^{-6} \cH^{\mu\nu}_{h\chiv(g_{\mu\nu},\chi)}\chiv\ ,
\\
\cH^{\rho\sigma}_{\chiv h(\Omega^2 g_{\mu\nu},\Omega^{-1}\chi)}(\Omega^2 h_{\rho\sigma})
=&\ \Omega^{-3} \cH^{\rho\sigma}_{\chiv h(g_{\mu\nu},\chi)}h_{\rho\sigma}\ ,
\\
\cH_{\chiv\chiv(\Omega^2 g_{\mu\nu},\Omega^{-1}\chi)}(\Omega^{-1}\chiv)
=&\ \Omega^{-3} \cH_{\chiv\chiv(g_{\mu\nu},\chi)}\chiv\ .
\end{align}
These transformation properties, which would have been hard to check in the form 
\eq{exp1}-\eq{exp4}, guarantee that the linearized action is invariant under the
``background Weyl transformations''
\be
g_{\mu\nu}\mapsto \Omega^2 g_{\mu\nu}\ ;\qquad
\chi\mapsto \Omega^{-1}\chi\ ;\qquad
h_{\mu\nu}\mapsto \Omega^2 h_{\mu\nu}\ ;\qquad
\chiv\mapsto \Omega^{-1}\chiv\ .
\ee
These parallel the familiar invariance under ``background diffeomorphisms''
which holds for any $\xi$. 

The Hessian can be regarded as a differential operator mapping
the covariant tensor $h_{\rho\sigma}$ to a contravariant tensor.
The trace and determinant of such an operator are basis dependent. 
In the quantum theory one needs a differential operator 
mapping covariant tensors to covariant tensors.
If we think of $(h_{\mu\nu},\chiv)$ as a vector in field space,
this corresponds to ``lowering the first index'' on the Hessian,
and is thus achieved by means of a metric in field space.
We choose the conformally invariant functional metric
\be
\cG((h_1,\chiv_1),(h_2,\chiv_2))=
\int d^4x\sqrt{g}\left[\chi^4 h_{1\mu\nu}g^{\mu\rho}g^{\nu\sigma}h_{2\rho\sigma}
+\chi^2\chiv_1\chiv_2\right]\ .
\ee
Then, the linearized action can be written in abridged notation as
\be
S^{(2)}=\frac{1}{2}Z\cH(\theta,\theta)=\frac{1}{2}Z\cG(\theta,\cO \theta)\ ,
\ee
where $\theta^T=\left(h,\chiv\right)$
and the components of $\cO$ are given by
\begin{align}
(\cO_{hh})_{\mu\nu}{}^{\rho\sigma}
&=
\chi^{-4}g_{\mu\alpha}g_{\nu\beta}\cH_{hh}^{\alpha\beta\rho\sigma}\ ,
\\
(\cO_{h\chiv})_{\mu\nu}
&=
\chi^{-4}g_{\mu\alpha}g_{\nu\beta}\cH_{h\chiv}^{\alpha\beta}\ ,
\\
\cO_{\chiv h}^{\rho\sigma}
&=
\chi^{-2}\cH_{\chiv h}^{\rho\sigma}\ ,
\\
\cO_{\chiv\chiv}
&=
\chi^{-2}\cH_{\chiv\chiv}\ .
\end{align}
From the symmetry of $\cH$ in its arguments, there follows self-adjointness
of $\cO$. Furthermore, the operator $\cO$ is Weyl-covariant in the sense that
\begin{align}
\label{henry}
(\cO_{hh(\Omega^2 g_{\mu\nu},\Omega^{-1}\chi)})_{\mu\nu}{}^{\rho\sigma}(\Omega^2 h_{\rho\sigma})
&=\Omega^2 (\cO_{hh(g_{\mu\nu},\chi)})_{\mu\nu}{}^{\rho\sigma}h_{\rho\sigma}\ ,
\\
(\cO_{h\chiv(\Omega^2 g_{\mu\nu},\Omega^{-1}\chi)})_{\mu\nu}(\Omega^{-1}\chiv)
&=\Omega^2 (\cO_{h\chiv(g_{\mu\nu},\chi)})_{\mu\nu}\chiv\ ,
\\
(\cO_{\chiv h(\Omega^2 g_{\mu\nu},\Omega^{-1}\chi)})^{\rho\sigma}(\Omega^2 h_{\rho\sigma})
&=\Omega^{-1} (\cO_{\chiv h(g_{\mu\nu},\chi)})^{\rho\sigma}h_{\rho\sigma}\ ,
\\
\cO_{\chiv\chiv(\Omega^2 g_{\mu\nu},\Omega^{-1}\chi)}(\Omega^{-1}\chiv)
&=\Omega^{-1} \cO_{\chiv\chiv(g_{\mu\nu},\chi)}\chiv\ .
\end{align}
Note that $\cO_{hh}$ and $\cO_{\chiv\chiv}$ are dimensionless. 
In particular, their leading terms begin with $-(1/\chi^2)D^2$.


\section{Gauge fixing}


Now we have to gauge fix for diffeomorphisms.
We add to the action the gauge fixing term
\be
S_{GF} =  \frac{1}{2\alpha}\int d^4x 
\sqrt{g}\,\frac{1}{2}Z\xi\chi^2 F_\mu {\bar g}^{\mu\nu} F_\nu\ ,
\ee
where
\be
\label{gf}
F_\nu =D_\mu h^\mu{}_\nu -\frac{\beta + 1}{4}D_\nu h
\ .
\ee
Note the appearance of the factor $Z\xi\chi^2/2$ which corresponds to the
usual factor of $1/16\pi G$ in the gravitational gauge fixing. 
It is there for dimensional reasons ($F_\mu$ is dimensionless) and allows the gauge fixing term to combine seamlessly with the inverse propagator \eq{exp5}-\eq{exp8}.
Integrating by parts and symmetrizing we have
\be
S_{GF}\! =\!  \frac{Z\xi}{4\alpha}\! \int\! d^4 x \sqrt {g}
\chi^2\Biggl[-h_{\mu\nu} D^\mu D_\rho h^{\rho\nu}
+\frac{1+\beta}{4}(h D^\mu D^\nu h_{\mu\nu}
+h_{\mu\nu} D^\mu D^\nu h)
-\frac{(1+\beta)^2}{16}hD^2 h
\Biggr].
\ee
The ghost action corresponding to the gauge \eq{gf} is given by
\begin{align}
S_{gh} &= \int d^4x \sqrt{g}\, \chi^2{\bar C}_\mu g^{\mu\nu} (\cO_{gh})_\nu^\rho C_\rho
\nonumber
\\
&=\ \cG_{gh}\left(\bar C,\cO_{gh} C\right)\ .
\label{gh}
\end{align}
where $\bar C$ and $C$ are dimensionless anticommuting vector fields,
\be
\cG_{gh}\left(A,B\right)=
\int d^4x \sqrt{g}\,\chi^2 A_\mu g^{\mu\nu}B_\nu
\ee
is the Weyl invariant inner product on vector fields and
\be
(\cO_{gh})_\mu^\nu=
-\frac{1}{\chi^2}\left(\delta_\mu^\nu D^2 +\frac{1-\beta}{2} D_\mu D^\nu +\cR_\mu{}^\nu\right)
\ee
is the Weyl-covariant operator acting on ghosts.

In the case $\xi=1/6$ the quadratic action also has zero modes \eq{gaugeweyl} 
corresponding to infinitesimal Weyl transformations. This also requires gauge fixing.
The most convenient way to fix this gauge is to choose $\chiv=0$.
With this choice we remain with the field $h_{\mu\nu}$
propagating in the background $g_{\mu\nu}$ and $\chi$.
This will allow us to immediately use results obtained previously for pure gravity.
With other gauge choices the dilaton fluctuations will remain,
and will generally mix with the graviton.
In these gauges the calculations will resemble more those in \cite{perini2,narain}.

We will choose the de Donder-Feynman gauge $\beta=1$, $\alpha=1$ which simplifies
the quadratic action considerably.
\footnote{It may be preferable to use the gauge $\beta=0$ which
only imposes conditions on the traceless part of $h_{\mu\nu}$
and therefore decouples the diffeomorphism from the Weyl gauge fixing.
This, however, would complicate the algebra without leading
to significant new insight.}
In this case the full gauge fixed quadratic action is simply
\be
S^{(2)}+S_{GF}=
\frac{1}{2} Z\int d^4 x \sqrt {g}\,
h_{\mu\nu}\cH^{\mu\nu\rho\sigma}h_{\rho\sigma}
\ee
with
\be
\label{qop}
\cH^{\mu\nu\rho\sigma}\!=\!
\frac{1}{12}\chi^2\Biggl[
K^{\mu\nu\rho\sigma} \!
\left(-D^2+\cR-12\lambda Z\chi^2\right)
-\cR^{\mu\rho\nu\sigma}\!
-g^{(\mu|\rho}\cR^{|\nu)\sigma}
+\frac{1}{2} (g^{\mu\nu}\cR^{\rho\sigma}\!
+\cR^{\mu\nu}g^{\rho\sigma})
\Biggr].
\ee


\section{The Weyl invariant effective average action}


We now assume $\xi=1/6$.
In order to evaluate the beta functions we construct the
so-called Effective Average Action (EAA), which is the effective
action with a cutoff on the propagation of low momentum modes.
It can be defined by the functional integral
\be
e^{-W(j,\bar\phi)}=\int d\mu\, e^{-S(\phi)
-\Delta S_k(\phi,\bar\phi)-\int j\phi}
\ee
followed by Legendre transform: 
\be
\Gamma_k(\phi,\bar\phi)=
W_k(j,\bar\phi)
-\int j\phi
-\Delta S_k(\phi,\bar\phi)\ .
\ee
Here $\phi$ denotes collectively the quantum fields, $\bar\phi$
the backgrounds, $j$ the sources and
$\Delta S_k$ is the cutoff action, to be specified shortly.
We shall be interested in the functional $\Gamma_k(\bar\phi)=\Gamma_k(0,\bar\phi)$
where the expectation values of the fluctuations are set to zero.

It is sometimes assumed that when a classical theory with action $S$
is scale invariant, the appearance of a cutoff and/or renormalization scale
in the definition of the functional integral 
{\it necessarily} leads to a breaking of scale invariance.
A fortiori, this is expected for Weyl invariance.
This is however not the case in presence of a dilaton field.
This point has been made early on, in the context of dimensional regularization, in \cite{englert}.
The existence of a choice in the quantization procedure has been discussed in \cite{flop}.
This choice has been discussed in the context of the RG flow of conformally reduced gravity in \cite{creh}.
The same point has also been reiterated more recently in \cite{shapo},
where it was shown how to maintain Weyl invariance in the presence of a lattice cutoff.
In the following we shall see how the cutoff $\Delta S_k$ can be defined
in such a way as to preserve Weyl invariance.

First, we assume that the measure is formally Weyl invariant.
This can be achieved by writing it in the form
\be
d\mu=
\prod (\chi^2dh_{\mu\nu})\prod\left(\frac{d\eta}{\chi}\right)
\prod d\bar C_\mu\prod dC_\nu\ ,
\ee
where the product extends over all normal modes of the fields.
Note that the $h$, $\eta$ measures can be written simply
in terms of the Weyl-invariant (dimensionless) variables
$\hat h_{\mu\nu}=\chi^2 h_{\mu\nu}$, $\hat\eta=\eta/\chi$.

The first step in the definition of the cutoff
is to choose a differential operator
whose eigenfunctions we will take as a basis in field space,
and in terms of whose eigenvalues the cutoff will be imposed.
Since we want to define the cutoff in such a way as to preserve
the background Weyl invariance, the operator has to satisfy
the covariance property 
\be
\label{covariance}
\Delta_{(\Omega^2 g_{\mu\nu},\Omega^{-1}\chi)} \left(\Omega^2 h_{\mu\nu}\right)
=\Omega^2 \Delta_{(g_{\mu\nu},\chi)} h_{\mu\nu}\ .
\ee
This property has the following important consequence:
if $h_{\mu\nu}$ is an eigenfunction of $\Delta_{(g_{\mu\nu},\chi)}$
with eigenvalue $\lambda$, then $\Omega^2 h_{\mu\nu}$
is an eigenfunction of $\Delta_{(\Omega^2 g_{\mu\nu},\Omega^{-1}\chi)}$
with the same eigenvalue.
Therefore, the spectrum of $\Delta$ is Weyl invariant.

We then have to introduce a cutoff scale $k$.
By definition it must have dimension of mass and therefore
it must transform under Weyl transformations as $k\to\Omega^{-1}k$.
In general, it is therefore a function on spacetime.
The ratio $u=k/\chi$ is a dimensionless and hence Weyl-invariant function.
In the following we will assume that $\chi$ and $k$ are proportional,
so that $u$ is constant.
Then, recalling that the eigenvalues of $\Delta$ are dimensionless,
we can use this number $u$ as the cutoff on the spectrum of $\Delta$.

Concretely, we proceed as follows.
We assume that $\Delta=\cO_{hh}$ for the graviton sector and
$\Delta=\cO_{gh}$ for the ghost sector.
These operators satisfy the desired Weyl covariance properties
(see \eq{henry}).
In the gauge that we have chosen above, where there is no $\eta$ fluctuation,
we define the cutoff action by
\bea
\label{cutoffaction}
\Delta S_k
&=&\frac{1}{2}Z\cG\left(h,\frac{1}{12}\frac{1}{\chi^2}R_k(\chi^2\cO_{hh})h\right)
+\cG_{gh}\left(\bar C,\frac{1}{\chi^2}R_k(\chi^2\cO_{gh})C\right)
\ ,
\eea
with 
\be
\label{cutf}
\frac{1}{\chi^2}R_k(\chi^2\cO)
=\frac{k^2}{\chi^2}\,r\!\left(\frac{\chi^2}{k^2}\cO\right)\ .
\ee

In order to define a cutoff, the function $r(y)$ must be monotonic, 
it must tend rapidly to zero for
$y>1$ and can be normalized so that $r(0)=1$.
We shall use the optimized cutoff \cite{optimized} 
\be
\label{optimized}
r(y)=(a-y)\theta(a-y)
\ee
where $\theta$ is the Heaviside step function and $a$ is a free parameter
that corresponds to choosing different renormalization schemes.
In the following it will be set to one unless otherwise stated. 

The crucial fact to notice at this point is that the cutoff action
\eq{cutoffaction} is invariant under Weyl transformations.
Normally a cutoff is regarded as a fixed scale breaking Weyl invariance,
but with our construction every appearance of the cutoff is neutralized
by a compensating factor of the dilaton:
the EAA depends on the cutoff only through the dimensionless,
Weyl-invariant, constant combination $u$.
Since the full quadratic action, including the cutoff, is background Weyl invariant, 
and the functional measure is also background Weyl invariant, the EAA is background Weyl invariant:
\be
\label{quantumweylinvariance}
\Gamma_{\Omega^{-1}k}(\Omega^2g_{\mu\nu},\Omega^{-1}\chi)
=\Gamma_k(g_{\mu\nu},\chi)\ .
\ee

In the context that we are considering here,
with a cutoff that could in general be a function on spacetime,
the ``beta functional'' of the theory is defined as
\be
\label{bfnl}
\int dx\,k(x)\frac{\delta\Gamma_k}{\delta k(x)}\ .
\ee
It is always possible, by means of a Weyl transformation,
to go to a conformal frame where the cutoff is constant,
and in this frame the beta functional reduces to
\be
k\frac{d\Gamma_k}{d k}\ .
\ee
Quite generally, given that the EAA depends on $k$ only through the
constant $u$, we have
\be
\int dx\,k(x)\frac{\delta\Gamma_k}{\delta k(x)}
=u\frac{d\Gamma_k}{d u}\ .
\ee

\section{Beta functions}

Since our purpose here is to illustrate the properties of
the Weyl invariant theory, and not to obtain accurate estimates 
for the beta functions, for the sake of simplicity we shall restrict
our attention to the one loop approximation.
The one loop EAA is
\be
\label{effac}
\Gamma_k^{(1)}(g_{\mu\nu},\chi)=S(g_{\mu\nu},\chi)
+\frac{1}{2}\mathrm{Tr}\log\left(\frac{P_k(\chi^2\cO_{hh})}{\chi^2}\right)
-\mathrm{Tr}\log\left(\frac{P_k(\chi^2\cO_{gh})}{\chi^2}\right)\ ,
\ee
where $P_k(z)=z+R_k(z)$.
The only $k$-dependence is in the cutoff functions and one gets:
\begin{align}
\label{ERGE}
u\frac{d\Gamma_k}{d u}&=
\frac{1}{2}\mathrm{Tr}
\frac{\chi^2}{P_k(\chi^2\cO_{hh})}u\frac{d}{du}\left(u^2 r\left(\frac{\cO_{hh}}{u^2}\right)\right)
-\mathrm{Tr}
\frac{\chi^2}{P_k(\chi^2\cO_{gh})}u\frac{d}{du}\left(u^2 r\left(\frac{\cO_{gh}}{u^2}\right)\right)
\nonumber
\\
&=\mathrm{Tr}\,\theta\left(u-\cO_{hh}\right)
-2\mathrm{Tr}\,\theta\left(u-\cO_{gh}\right)\ ,
\end{align}
where we have used equation \eq{cutf}, and then 
the explicit form of the optimized cutoff \eq{optimized}.
Here we see explicitly the Weyl invariance of the beta functional.
Note that unlike the EAA itself, its $u$-derivative is UV finite
and therefore there cannot be any issue of hidden breaking
of Weyl invariance due to some UV regulator.
If the initial point of the flow is chosen to be Weyl invariant,
it will remain so for all $u$.

The standard way of extracting beta functions from this expression, is to make an ansatz of the form
\be
\label{localexp}
\Gamma_k=\sum_i g_i \int d^4x\,\sqrt{g}\,\cM_i
\ee
and to extract from the r.h.s. of \eq{ERGE} the coefficients of the operators $\cM_i$.
This method was first applied to gravity in \cite{reuter1}.
Here we will truncate the expansion to the first two terms.
The evaluation of the traces has been done many times before,
and we refer for example to \cite{cpr2}.
In a gauge in which $k$ and $\chi$ are constant,
the first two terms of an expansion in derivatives of the metric are
\be
\label{betafnctnl}
k\frac{d\Gamma_k^{(1)}}{dk} = 
\frac{1}{16\pi^2}k^4
\int d^4 x \sqrt{g}
-\frac{1}{16\pi}\frac{23}{3\pi}k^2
\int d^4 x \sqrt{g} R
\ .
\ee

\subsection{The Einstein approach: treating $\chi$ as coupling}

The traditional procedure is to read from \eq{betafnctnl} the
beta functions of the dimensionless couplings
$\tilde\Lambda=\Lambda/k^2$, $\tilde G=G k^2$,
in a gauge in which $k$ is constant.
One truncates the expansion \eq{localexp} to the first two terms,
in such a way that the EAA has the form \eq{hilbert}, or equivalently
\be
\label{sandra}
\int d^4x\sqrt{g}\frac{1}{16\pi\tilde G}
\left(2\tilde\Lambda k^4-k^2 R\right)\ .
\ee
Deriving this equation with respect to $k$ and comparing with 
\eq{betafnctnl} one finds:
\bea
\label{betaG}
k\frac{d\tilde G}{dk}&=&2\tilde G-\frac{23}{3\pi}\tilde G^2\ ,
\\
\label{betaLambda}
k\frac{d\tilde\Lambda}{dk}&=&
-2\tilde\Lambda
-\frac{23}{3\pi}\tilde G\tilde\Lambda
+\frac{1}{2\pi}\tilde G\ .
\eea
These beta functions have a FP at 
\be
\label{fp1}
\tilde G=\frac{6\pi}{23}\ ;\qquad
\tilde\Lambda=\frac{3}{92}\ .
\ee
Note that this procedure is equivalent to treating $\chi$ as a coupling.
More precisely, defining $\tilde\chi=\chi/k$ and using the relations
\be
\label{relat}
Z\tilde\chi^2=\frac{12}{16\pi \tilde G}\ ;\qquad
\lambda=\frac{2\pi}{9}\tilde\Lambda\tilde G\ ,
\ee
we see that these couplings have a FP at 
\be
\label{fp2}
(Z\tilde\chi^2)_*=\frac{23}{8\pi^2}\ ;\qquad
\lambda_*=\frac{\pi^2}{529}\ .
\ee
Since the main point of the Weyl invariant reformulation of gravity
was precisely to replace Newton's constant (and all other dimensionful couplings) 
by dimensionless couplings times powers of a field,
this interpretation is contrary to the philosophy we adopted in this paper.
Nevertheless, we would like to preserve the equivalence between the
Weyl invariant theory and Einstein gravity also at the quantum level.
We will see below how this is achieved.

\subsection{The Weyl approach: treating $\chi$ as field}

In the spirit of Weyl's theory, we interpret the dilaton
as a field providing an arbitrary local unit of mass.
It can be an arbitrary positive function.
The cutoff is assumed to be 
constant when measured in units of the dilaton.
In other words, the cutoff is parameterized by the dimensionless, 
Weyl-invariant real parameter $u=k/\chi$.
The dimensionless couplings $Z$ and $\lambda$,
in \eq{confaction} should be thought of as functions of $u$.

Equation \eq{ERGE} shows explicitly that the beta functional of the theory,
which gives the dependence of the EAA on $u$, is Weyl invariant.
Therefore, equation \eq{betafnctnl} can be regarded as a special case of equation
\be
\label{betafnctnl2}
u\frac{d\Gamma_k^{(1)}}{du} = 
\frac{1}{16\pi^2}u^4
\int d^4 x \sqrt{g}\chi^4
-\frac{1}{16\pi}\frac{23}{3\pi}u^2
\int d^4 x \sqrt{g}\chi^2 \cR
\ ,
\ee
written in a gauge in which $k$ and $\chi$ are constant.
Comparing \eq{betafnctnl2} with the $u$-derivative of \eq{confaction}
\be
u\frac{d\Gamma_k}{du}=
u\frac{d(Z^2\lambda)}{du}\int d^4x\sqrt{g}\chi^4
-\frac{1}{12}u\frac{d Z}{du}\int d^4x\sqrt{g}\chi^2 \cR
\ee
we obtain the beta functions:
\bea
\label{betaZ1}
u\frac{d Z}{du}&=&\frac{23}{4\pi^2}u^2\ ,
\\
\label{betalambda1}
u\frac{d\lambda}{du}&=&\frac{u^2-184 Z\lambda}{16\pi^2 Z^2}u^2\ .
\eea
These beta functions do not have a FP for $Z$ and $\lambda$,
in fact even the question of the existence of a FP is ill posed
because of the explicit appearance of the independent variable $u$
in the flow equations.
However, these equations are sufficiently simple that one can obtain 
their general solution.
The equation for $Z$ is independent of $\lambda$ and has the solution
\be
\label{Zsol}
Z(u)=Z_0+\frac{23}{8\pi^2}u^2
\ee
where $Z_0=Z(0)$ is the IR limiting value of $Z$.
Plugging this solution in the equation for $\lambda$ one finds
\be
\label{lamsol}
\lambda(u)=
\frac{\pi^2(u^4+64\pi^2Z_0^2\lambda_0)}{(8\pi^2 Z_0+23 u^2)^2}
\ee
where $\lambda_0=\lambda(0)$ is the IR limiting value of $\lambda$.

In this interpretation, the trajectory depends on the arbitrary integration
constants $Z_0$ and $\lambda_0$, but for $u\to\infty$ all
trajectories have the same asymptotic behavior:
$Z$ grows asymptotically like $\frac{23}{8\pi^2}u^2$
and $\lambda$ tends to the limiting value $\pi^2/529$.
This agrees with the results \eq{fp2} in the following way:
the limiting value of $\lambda$ is the fixed point value,
and the asymptotic value of $Z/u^2=Z\tilde\chi^2$ is also the fixed point value.

We observe that the trajectory reaches the FP for $u\approx Z_0$.
There is a trajectory that corresponds to the theory sitting forever at the FP: 
it is the limiting trajectory characterized by $Z_0=0$, with $\lambda_0$ arbitrary.
Note that this is not inconsistent with the requirement $Z>0$,
because for any finite $u$ this condition will be satisfied.


\section{Discussion}


\subsection{$Z$ vs. $\xi$}

Since the flow equation preserves Weyl invariance,
the condition $\xi=1/6$ that we assumed in section 6 will be maintained, independent of $k$.
We can say that $1/6$ is a FP of $\xi$.
We can now see why it was necessary to retain the redundant 
wave function renormalization constant $Z$ in the action \eq{action}.
Had we set it equal to one from the beginning, we might have been led to interpret 
the term proportional to $\cR$ in \eq{betafnctnl2} as a renormalization of $\xi$,
and we would have concluded that $\xi$ has a FP
at $23/8\pi^2$ instead of $1/6$.
Instead, $\xi$ should be seen as the relative weight between the 
second and the third term in \eq{action}, and $Z$ as the overall weight.
Then, knowing that $\xi=1/6$ we can read off the beta function of $Z$
from the $\chi^2 R$ term.
It would be nice to verify this result explicitly by reading off the
beta function of $Z$ from the term proportional to $(\nabla\chi)^2/2$, 
and checking that it gives the same beta function. 
We leave this check for the future.

A related observation is that since $Z$ is redundant (or ``inessential'')
the existence of a FP does not require it to reach a finite limit in the UV.
Instead, one would expect it to scale at a FP with some calculable power of momentum.
This is exactly what happens in the Weyl approach, see equation \eq{Zsol}.
Instead, it should come as a surprise that in the Einstein approach
all couplings seem to be fixed in the UV.

This has been discussed earlier in \cite{perini3}, where it
was attributed to the special role of the metric.
The Weyl-invariant reformulation of the theory casts some 
additional light on this issue.
In the Einstein approach one uses the cutoff $k$ as reference mass scale 
and there is no independent dilaton field.
In this treatment one is not free to absorb $Z$ in a redefinition
of the metric, because any such rescaling would affect also the cutoff,
which is supposed to be left untouched.
Since $Z$ cannot be scaled away, it must reach a finite limit at a FP.
In the Weyl approach one uses the dilaton field $\chi$ as a unit of mass
and the parameter $Z$ can be absorbed in a redefinition of $\chi$
without touching the cutoff.
In this case $Z$ is not required to have a finite limit,
and indeed it is found to scale to infinity.
In this approach, the resulting flow equations
\eq{betaZ1}, \eq{betalambda1} are not autonomous.
This is reminiscent of the observation that the flow equations
of ordinary Einstein-Hilbert gravity are not autonomous
when one uses Planck units \cite{perini3}.
Such equations don't have FPs, but asymptotically
$\lambda$ tends to a finite value, while $Z$ does not,
in accordance with their status of essential and inessential
parameter, respectively.
Though superficially different, the physics contained in the two
procedures discussed above is completely equivalent.

\subsection{Renormalization scheme dependence}

For simplicity in the preceding discussion we have set the
free parameter $a$ of equation \eq{optimized} to one.
It is interesting to discuss the dependence of the results on $a$.
If we allow $a\not=1$, then equations \eq{betaZ1}, \eq{betalambda1} 
get modified by the replacement of $u$ by $au$.
Then, the asymptotic behavior of $Z$ is modified by a factor $a^2$,
but the limiting value of $\lambda$ is independent of $a$.
In view of equation \eq{relat}, this accords with earlier observations regarding the 
scheme-independence of the dimensionless product $\Lambda G$ \cite{dou,lauscher}.
It also suggests that the FP value of $\tilde G$,
which in our calculation turned out to be
\be
\tilde G=\frac{12k^2}{16\pi Z\chi^2}=\frac{1}{a^2}\frac{6\pi}{23}\ ,
\ee
is really arbitrary as long as it is nonzero.
In particular we notice that if we want to take $k$
at a very low scale, we can guarantee that the asymptotic expansion
of the heat kernel, which is used in the derivation of \eq{betafnctnl},
is still valid by taking $a$ large.
This produces a FP for the dimensionless Newton's
constant which is small.

\subsection{Other cutoff schemes}

The calculation of the beta functions presented above was based
on the use of the operator $\cO$ in the definition of the cutoff.
This was called a ``type III'' cutoff in \cite{cpr2}.
It is simple to deal with in the gauge $\alpha=1$, because then the operator
$\cO$ is minimal.
For other gauges nonminimal terms would appear and 
it becomes more convenient to adopt the scheme IB, which 
is based on the decomposition of the graviton field $h_{\mu\nu}$ 
into its irreducible components.
The operator $\cO$ then decomposes, in any gauge, into (nearly) diagonal blocks,
acting separately on the spin 2,1 and 0 components of $h_{\mu\nu}$.
Then one can impose the cutoff separately in each block.

In the Weyl invariant case the procedure goes through nearly unchanged.
The decomposition of the graviton can be performed in a background Weyl invariant
way just replacing ordinary covariant derivatives $\nabla$ by Weyl covariant
derivatives $D$. The operators acting on the irreducible components are
then of the form $-D^2+U$. We can then choose the cutoff to be a kernel
constructed with the operator $-D^2$.
This operator satisfies the covariance property \eq{covariance}
and therefore the resulting effective average action is again
background Weyl invariant.
The only difference would be in the numerical value of the constants
that appear in the beta functions.

\subsection{No diffeomorphism anomalies}

One can now see that in the formalism presented above the cutoff
can be treated as a function on spacetime, without incurring in diffeomorphism anomalies.
This is basically due to the fact that in the Weyl-invariant theory all the couplings 
are dimensionless and constant over spacetime.
What might have been a position-dependence of the couplings has been transferred
to a position dependence of the dilaton.
With the procedure described in section 1.2, 
if the cutoff is some function $k(x)$, all the couplings
become functions $g_i(k(x))$, where the functional dependence $g_i(k)$
is given by the renormalization group flow
and the functional form of $k(x)$ depends on the cutoff choice.
This applies both to dimensionful and dimensionless couplings.
On the other hand with the procedure described in this paper, 
a dimensionful coupling $g_i$
is replaced first by $\hat g_i \chi^{d_i}$, which can
then be rewritten as $\hat g_i (\chi/k)^{d_i}k^{d_i}=\tilde g_i k^{d_i}$.
The factor $k^{d_i}=\chi^{d_i}u^{d_i}$ can be any positive function, but the
dimensionless combinations $\hat g_i$ and $\tilde g_i$ are constant.
In particular, couplings $g_i$ that are dimensionless to begin with,
such as the fine structure constant, remain always constant.
In this way the issues described in section 1.2 do not arise.

\subsection{Matter}

One can extend the preceding treatment to gravity coupled to any matter field,
with any kind of coupling.
By means of the dilaton one can convert all dimensionful parameters
in the matter action to dimensionless ones, and using the Weyl covariant derivatives
rewrite the action in a form that is manifestly Weyl invariant.

It is interesting to observe that the standard model is 
almost entirely Weyl invariant, even without using the dilaton.
It contains only renormalizable, dimensionless coupling constants
and it is written in such a way that all the masses of elementary particles
derive from their interactions with the Higgs field.
The only dimensionful parameter in the standard model is the
mass term in the Higgs potential, so this is the only term in the action
that would require a dilaton to be made Weyl invariant.
There have been several studies along these lines recently \cite{shapo2,nicolai}.
One particularly intriguing speculation, which seems to have been 
brought up independently several times \cite{cheng} 
is that there is no Higgs potential;
in this case there is obviously also no Higgs mass term.
Expanding around a nonzero scalar VEV,
all the scalar fluctuations are massless and can be regarded as Goldstone bosons:
the angular modes of the Higgs doublet can be viewed as the Goldstone bosons
of $SU(2)_L\times U(1)$ and the radial mode of the Higgs doublet
(which is an $SU(2)\times U(1)$ singlet) can be identified with the dilaton.
Electroweak symmetry breaking would be due to the $SU(2)_L\times U(1)$
Goldstone bosons, whose kinetic term requires a prefactor with
dimension of mass squared.
In the spirit of the present work, this prefactor would be written
as $\chi^2$ times a dimensionless coupling.
Likewise, in the Yukawa couplings the scalar VEV would be replaced by $\chi$.
The mass of all fields would thus be proportional to $\chi$.
The phenomenology of such a model is probably very similar to that
of an asymptotically safe Higgsless standard model \cite{higgsless}.

It is easy to calculate the contribution of 
minimally coupled matter fields to
the beta functions of $Z$ and $\lambda$ \cite{perini1,largen}.
We will not pursue this further here.

\section{Outlook}

We have shown how to rewrite a theory of gravity in a Weyl invariant form,
and how to write renormalization group equations that preserve this property.
Fluctuations of the dilaton are unphysical and can be eliminated
by a gauge choice, while background Weyl invariance is preserved. 

The calculations presented here can be viewed as generalizations of the work
on conformally reduced gravity \cite{creh,machado}.
The difference lies in the fact that in those earlier references
only the conformal degree of freedom of the metric was allowed
to fluctuate, whereas here the full metric is dynamical.
The main result however is the same: Weyl invariance can be preserved.
An alternative quantization method exists, which does not preserve
Weyl invariance: it leads to the appearance of a nontrivial potential for
the dilaton \cite{flop}. As emphasized in \cite{creh},
this alternative quantization corresponds to 
the standard way of quantising a scalar field theory.

By recasting the theory in a Weyl invariant form, one trades all
dimensionful couplings for dimensionless ones.
All the scales that are so evident in the
low energy world would then be proportional to a single scale,
which is set by the background dilaton.
Although here we have decided to discuss only the Hilbert action,
it is clear that all the higher terms in the derivative expansion of
the gravitational action can be treated in the same way.
Furthermore, the weak and the QCD scale can also emerge in a similar fashion.
The VEV of the dilaton in itself is physically unmeasurable: only the ratios of
mass scales is physically meaningful.
They could, in principle, be computable \cite{thooft}.
We hope to return to these issues elsewhere.

The other motivation for this work was the formulation of
a quantum field theory with non-constant cutoff that is free of
the diffeomorphism anomalies discussed in the introduction.
This goal has been achieved by making all couplings dimensionless
functions of the constant, dimensionless parameter $u=k/\chi$.
The resulting theory is Weyl invariant and given any nonconstant
cutoff $k(x)$ one can always go to another gauge where $k$ is constant.
Of course, in this new gauge the metric will also be conformally transformed,
but the physics will be the same.
The generalization of the theory to the case
when the ratio $u$ is an arbitrary (dimensionless) function seems problematic,
but even within the limited freedom afforded by the present
formalism, it will be interesting to develop
applications to astrophysical and cosmological problems.

We conclude by mentioning two possible extensions of this work.
The first is to study the case when the abelian gauge field $\kappa_\mu$ is not flat.
Since the abelian gauge field behaves in many ways like
the electromagnetic field, the recent calculation of beta functions
of QED coupled to QEG should be a useful guide \cite{harst}.
The second is the extension of Weyl transformations \eq{conftransf}
to local $GL(4)$ transformations 
$g_{\mu\nu}\to \Omega^\rho{}_\mu\Omega^\sigma{}_\nu g_{\rho\sigma}$.
These transformations have been studied some times ago in a slightly different
context \cite{floreanini}. It would be interesting to reconsider
them in light of subsequent developments.

\bigskip
\bigskip
\goodbreak

\end{document}